\documentclass[letterpaper, 10 pt, conference]{ieeeconf}  

\IEEEoverridecommandlockouts                              

\usepackage{graphicx}
\usepackage{subcaption} 
\usepackage{xcolor}
\usepackage{multirow}
\usepackage{hyperref}
\usepackage{url}

\title{\LARGE \bf Traffic Reconstruction and Analysis of Natural Driving Behaviors at Unsignalized Intersections} 

\author{Supriya Sarker, Bibek Poudel, Michael Villarreal, Weizi Li \\ \texttt{\{ssarker8,bpoudel3,tvillarr\}@vols.utk.edu}, \texttt{weizili@utk.edu}}

\begin{document}

\maketitle
\thispagestyle{empty}
\pagestyle{empty}


\begin{abstract}

This paper explores the intricacies of traffic behavior at unsignalized intersections through the lens of a novel dataset, combining manual video data labeling and advanced traffic simulation in SUMO. This research involved recording traffic at various unsignalized intersections in Memphis, TN, during different times of the day. After manually labeling video data to capture specific variables, we reconstructed traffic scenarios in the SUMO simulation environment. The output data from these simulations offered a comprehensive analysis, including time-space diagrams for vehicle movement, travel time frequency distributions, and speed-position plots to identify bottleneck points. This approach enhances our understanding of traffic dynamics, providing crucial insights for effective traffic management and infrastructure improvements.

\end{abstract}

\section{INTRODUCTION} 
With the urban population expanding by approximately 60 million individuals annually, the mounting pressures on urban infrastructures are undeniable. This surge presents multifaceted challenges, with profound implications for environmental sustainability, economic vitality, and the overall quality of urban life. Consequently, urban planning must evolve to accommodate such an uptick, necessitating the development of transport systems and infrastructural strategies that cater to the burgeoning demands \cite{rapelli2019tust}. As urban areas swell, the incidence of traffic congestion is set to escalate correspondingly. The onus is on urban planners to craft and execute transport solutions and infrastructural frameworks adept at managing the present influx and anticipating future needs \cite{rapelli2021vehicular}. The pivotal role of traffic surveillance and management in shaping efficient transportation networks is unequivocally recognized by professionals in the field \cite{barmpounakis2020new}.

Simulation emerges as a pivotal tool in this landscape, with advancements in computational capabilities and diversified data sources enabling a granular representation of expansive urban territories. Historically utilized in the analysis of traffic patterns, simulations are increasingly employed in mobility studies, thanks to technological strides that enable the scrutiny of vast areas \cite{rapelli2019tust}. These analytical endeavors are critical for urban planners as they discern and evaluate the key conduits connecting residential zones with commercial hubs, determining the necessity for enhancements or the development of new routes \cite{rapelli2021vehicular}.
Traffic simulation stands as the preeminent instrument for predictive mobility analysis \cite{rapelli2021vehicular}. Utilizing vehicle trajectory data, which encapsulates the vehicular movement patterns, including location and time stamps, has been instrumental in a multitude of transportation domains. Applications range from traffic surveillance \cite{kong2018lotad}, volume inference~\cite{meng2017city}, and predictive modeling \cite{yin2021deep}, to dynamic forecasting \cite{li2023dynamic}, \cite{jin2022automated}, mobility analysis~\cite{Wang2021Mobility,Lin2021Safety}, and strategic traffic management~\cite{zheng2019learning,ma2020multi,ma2023mixed}.

Researchers are increasingly drawn to studying unsignalized traffic scenarios due to their prevalence in rapidly urbanizing areas and the need for cost-effective, efficient traffic management solutions. This focus is driven by the desire to enhance safety and optimize flow in these less structured environments, particularly as technological advancements like autonomous vehicles become more integrated into everyday traffic systems~\cite{villarreal2023can,villarreal2023hybrid,wang2023learning,Wang2023Privacy}. Since understanding unsignalized intersections is pivotal for devising flexible traffic strategies suitable for both urban and rural contexts, analyzing this type of scenario holds utmost significance.

In this project, we conducted a detailed recording of the traffic dynamics at key intersections in Memphis, TN, USA, capturing four hours of video during various peak periods including morning, midday, and afternoon. This visual data enabled us to manually extract pivotal information such as the timing, as well as the origin and destination lanes, of each vehicle navigating the intersection. The primary motivations of this paper are to present a thorough analysis of unsignalized intersections. Our research significantly contributes to this field by outlining the processes of data collection, extraction, and reconstruction; pursuing an in-depth examination of the traffic dynamics at these intersections, providing valuable insights that can guide improvements in road design and traffic management.





\section{RELATED WORK}
\label{related}
\subsection{Closed Course Testbeds}
Closed course testbeds refer to controlled environments specifically designed for testing and evaluating vehicles and transportation technologies. Unlike open road testbeds which are set in public roads, closed courses are private, restricted areas that simulate various road conditions and scenarios. This setup allows for precise control over testing conditions and is safer for experimental technologies like autonomous vehicles, as it eliminates interactions with regular traffic and uncontrolled variables. 
\subsection{Open Road Testbeds}

\section{Videos Collection System Description}
\label{system_description}
\subsection{Video 1}
\begin{itemize}
    \item Location: Goodlett Street and Walnut Grove Road
    \item Coordinate: (Latitude, Longitude) is (35.131508, -89.925530)
    \item Recording time of day: 5:00 PM to 6:00 PM
    \item Camera Position and Orientation: The camera is positioned at 10 ft height from the ground, at one of the corners of the intersection, pointing slightly downwards. 
\end{itemize}

\subsection{Video 2}
\begin{itemize}
    \item Location: Walnut Grove Road and Mendenhall Road
    \item Coordinate: (Latitude, Longitude) is (35.130825, -89.898503)
    \item Recording time of day: from 12:27 PM to 1:27 PM
    \item Camera Position and Orientation: GoPro was placed on St. Agnes School’s brick sign, facing towards the intersection. 
    \item Video Cameras Description and Processing: The video was shot using a GoPro Hero 9 at 30 FPS, 1080p, and using SuperView (essentially an ultra wide shot).
\end{itemize}

\subsection{Video 3}
\begin{itemize}
    \item Location: Walnut Grove Road and Mendenhall Road
    \item Coordinate: (Latitude, Longitude) is (35.130825, -89.898503)
    \item Recording time of day: Rush hour time period from 5:21 PM to 6:40 PM
    \item Camera Position and Orientation: GoPro was placed on St. Agnes School’s brick sign, facing towards the intersection.
    \item Video Cameras Description and Processing: The video was shot using a GoPro Hero 9 at 30 FPS, 1080p, and using SuperView (essentially an ultra wide shot).
\end{itemize}

\subsection{Video 4}
\begin{itemize}
    \item Location: Goodlett Street and Walnut Grove Road
    \item Coordinate: (Latitude, Longitude) is (35.131508, -89.925530)
    \item Recording time of day: 12:00 PM to 1:00 PM
\end{itemize}

\section{Reconstruction of Road Network}
\label{trajectory_data}

\subsection{Dataset Description}
The dataset, meticulously compiled from four hours of video recordings at key intersections in Memphis, TN, offers an in-depth look at the varying dynamics of traffic flow throughout different periods of the day, encompassing both the peak rush hours and quieter intervals. The data, which has been manually labeled and organized into four separate CSV files corresponding to each hour as detailed in the video collection system section, encompasses several critical attributes:

\begin{itemize}
\item Timestep: This represents the specific moment when the data was recorded, arranged in a continuous sequence that suggests a regular and systematic interval of data collection.
\item Start Lane: Identifies the lane from which a vehicle initiates its approach to the intersection, providing essential insights into the traffic flow patterns and pinpointing the vehicles' origins.
\item End Lane: Denotes the lane where the vehicle exits the intersection, which is instrumental in analyzing traffic directions and the intended destinations within the intersection's framework.
\end{itemize}

Each of the four CSV files is a representation of an hour's worth of data, offering a granular view of traffic behavior and patterns. The locations and names of the two intersections in Memphis city, where these recordings were captured, are detailed, offering context and specific geographical relevance to the dataset.

\subsection{Road Network Creation}  
In our project focused on analyzing traffic dynamics in Memphis, TN, we embarked on a journey to capture a real-world road network for our simulations. The first step in this endeavor involved sourcing accurate and detailed road network data. For this, we turned to OpenStreetMap (OSM), a comprehensive and open-source mapping service. We meticulously selected a specific area within Memphis, ensuring that it encapsulated the roads and intersections of interest. This careful selection was crucial for the fidelity of our simulation, as it would provide the groundwork upon which all further analysis would be built.

Once the area was defined, we extracted the corresponding map data from OSM. This data, rich with details about the roads, lanes, and intersections of Memphis, served as the raw material for our simulation model. The challenge then was to translate this raw map data into a format compatible with SUMO (Simulation of Urban MObility), our chosen traffic simulation software. To achieve this, we utilized a tool provided by SUMO called NETCONVERT. This powerful tool seamlessly transformed the OSM data into a .net.xml file, which is the native format understood by SUMO for representing road networks. In doing so, we were careful to maintain the integrity of the road characteristics as represented in the OSM data, ensuring that our simulation would reflect real-world conditions as closely as possible. The final step was to validate this newly created SUMO network file against the actual road layout of Memphis, a crucial process to ensure accuracy and reliability in our subsequent simulations. This thorough approach allowed us to lay a solid foundation for conducting meaningful and realistic traffic analysis using SUMO.

\subsection{Unsignalized Road Traffic Scenario Reconstruction}  
For our convenient of labeling we did assign a integer number to each lane as id. However. while creating network file, an id is assigned to a particular lane in SUMO's conventional form. During data processing, we map each integer lane\_id to SUMO assigned lane\_id so that we can reconstruct the same traffic pattern in each lane at a particular time. In the fig, we pointed the camera position that was recording the video and all the integer ids of each of the 4 edges. There are some internal lanes inside the intersection area that are assigned not by us; but selected by road network.  

\begin{figure*}[htbp]
\centering
\begin{subfigure}{0.30\textwidth}
   \includegraphics[width=\linewidth]{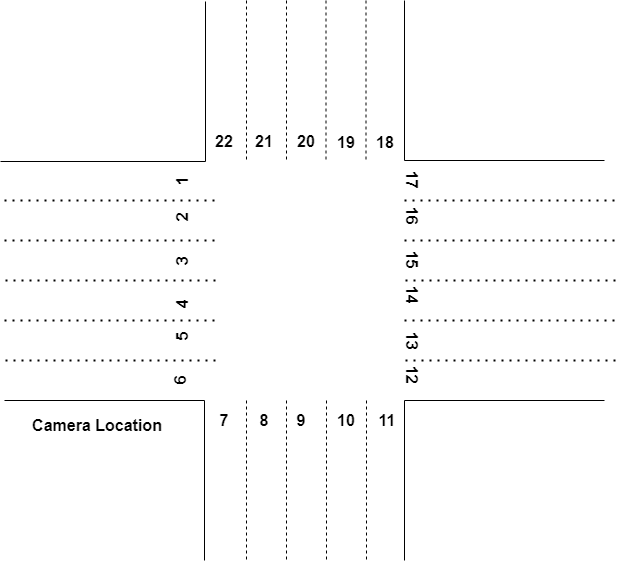}
   \caption{Video 1 lanes label}
   \label{fig:intersection_1_lanes_number}
\end{subfigure}%
\begin{subfigure}{0.30\textwidth}
   \includegraphics[width=\linewidth]{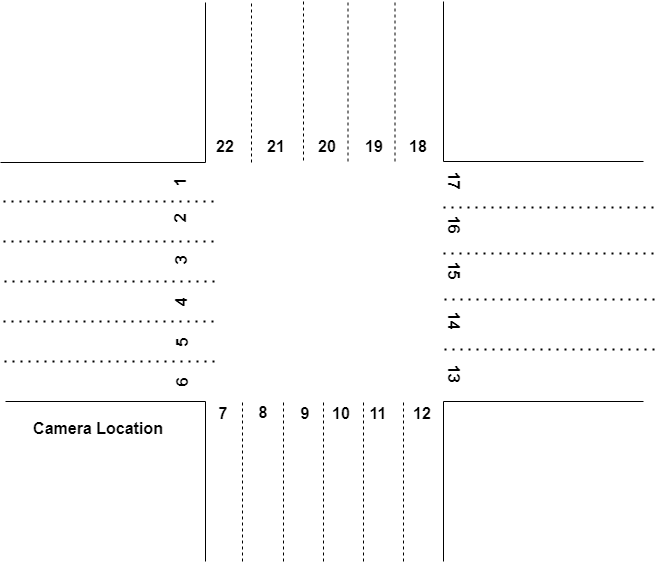}
   \caption{Video 2 and 3 lanes label}
   \label{fig:intersection_2_3_lanes_number}
\end{subfigure}%
\begin{subfigure}{0.30\textwidth}
   \includegraphics[width=\linewidth]{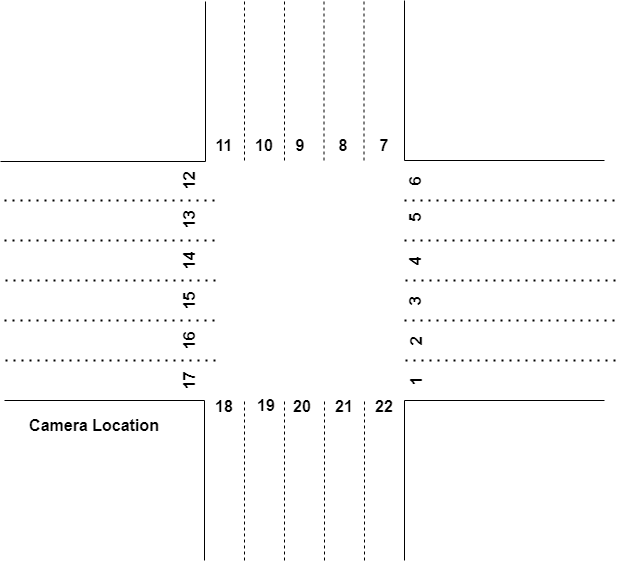}
   \caption{Video 4 lanes label}
   \label{fig:intersection_4_lanes_number}
\end{subfigure}

\caption{We have assigned lane ids to each lane for (a) Video 1, (b) Video 2 and 3, (c) Video 4, respectively. in the datasets the lanes are labeled with this number.}
\label{fig:lanes_number}
\end{figure*}

\begin{figure*}[htbp]
\centering
\begin{subfigure}{0.30\textwidth}
   \includegraphics[width=\linewidth, height=4cm]{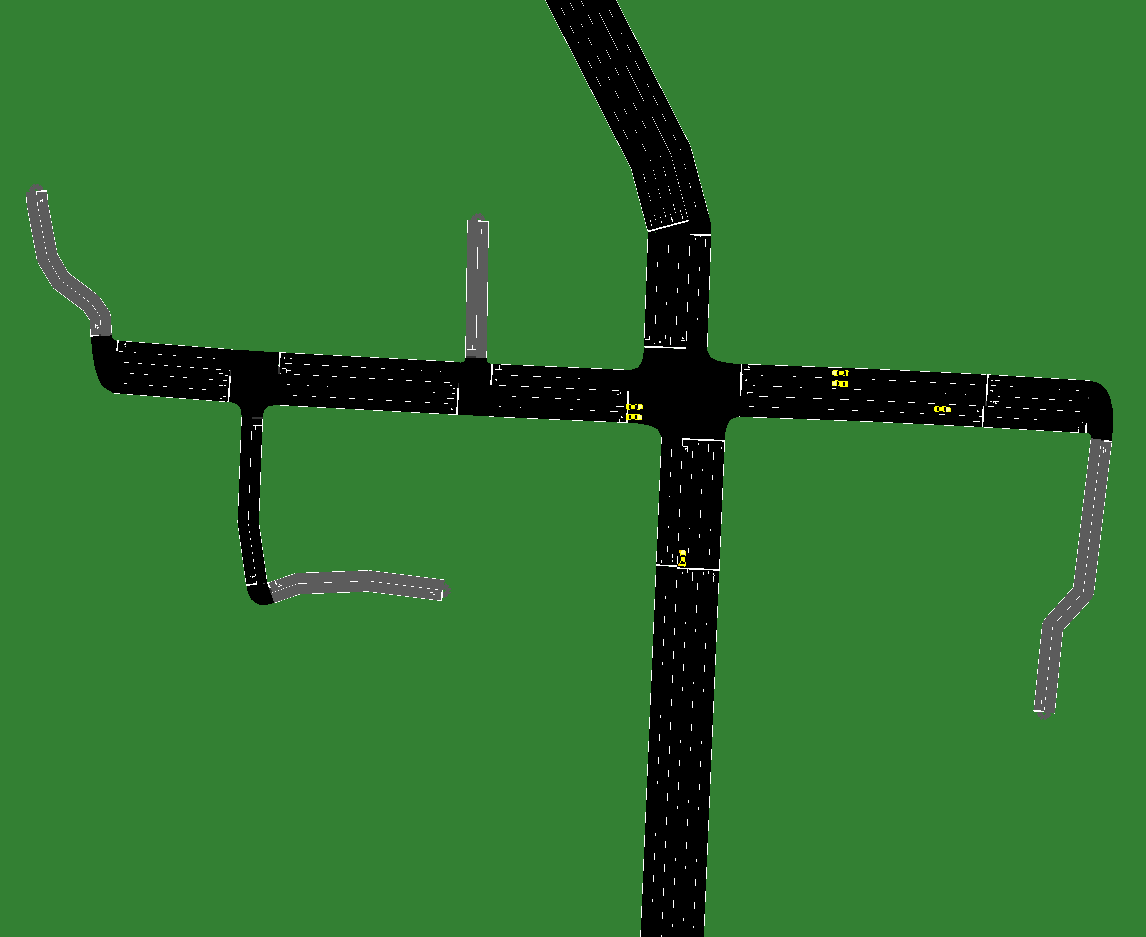} 
   \caption{Intersection of video 1 and 4 in SUMO simulation}
   \label{fig:intersection_2_4_sim}
\end{subfigure} \hspace{5pt}
\begin{subfigure}{0.30\textwidth}
   \includegraphics[width=\linewidth, height=4cm]{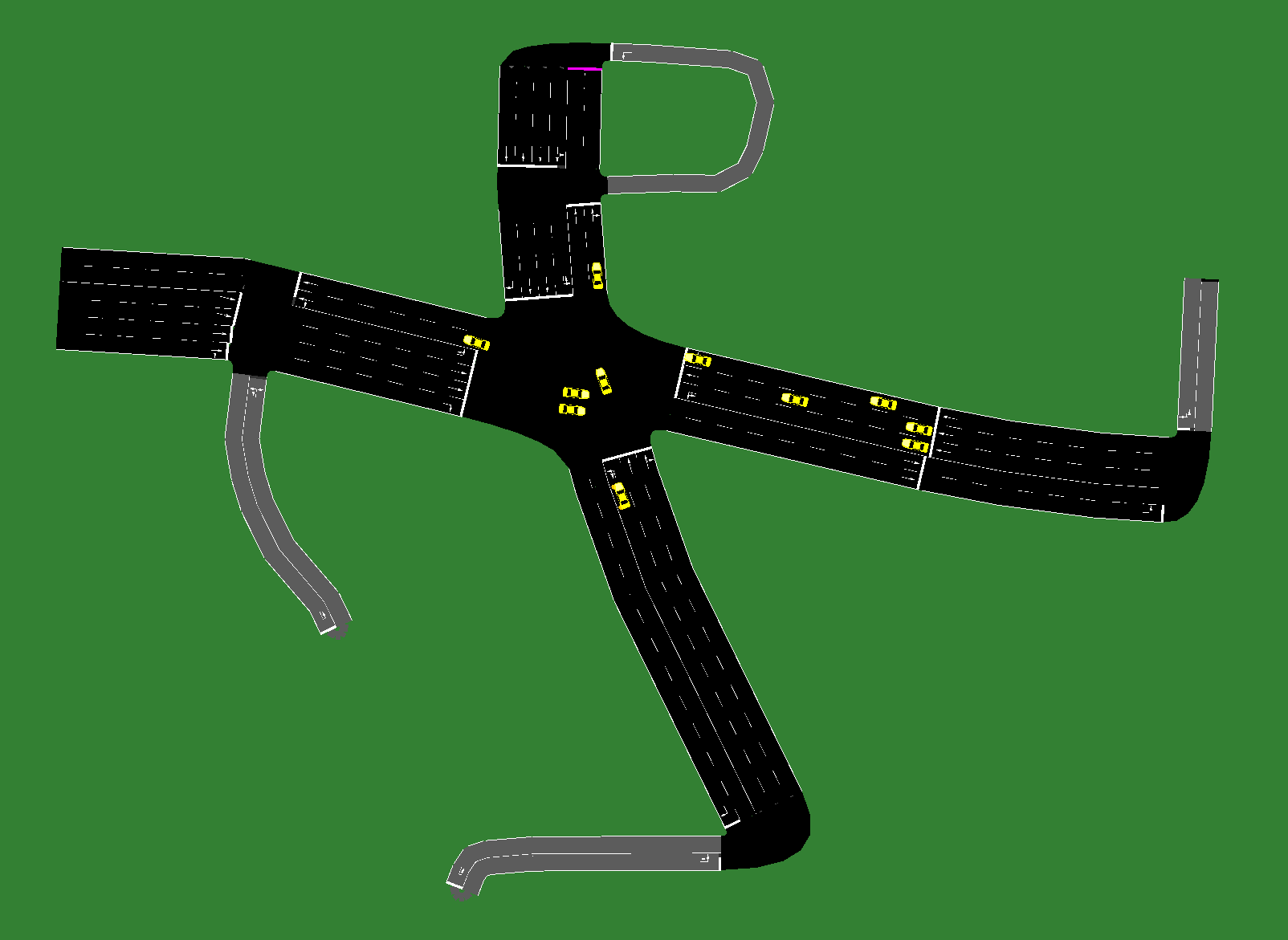} 
   \caption{Intersection of video 2 and 3 in SUMO simulation}
   \label{fig:intersection_2_3_sim}
\end{subfigure}%

\caption{We have assigned lane ids to each lane for (a) Video 1, (b) Video 2 and 3, (c) Video 4, respectively. in the datasets the lanes are labeled with this number.}
\label{fig:lanes_number}
\end{figure*}

\section{Intersection Analysis From Simulation}
\label{analysis}
\subsection{Time-Space Diagram}
In section \ref{analysis} we will discuss time-space diagram which is a visual representation of traffic flow through an intersection over time and space. The purpose of analysing time-space diagram is to analyze the behavior and efficiency of traffic movement through the intersection.  

\begin{figure*}[htbp]
\centering
\begin{subfigure}{0.24\textwidth}
   \includegraphics[width=\linewidth, height=4cm]{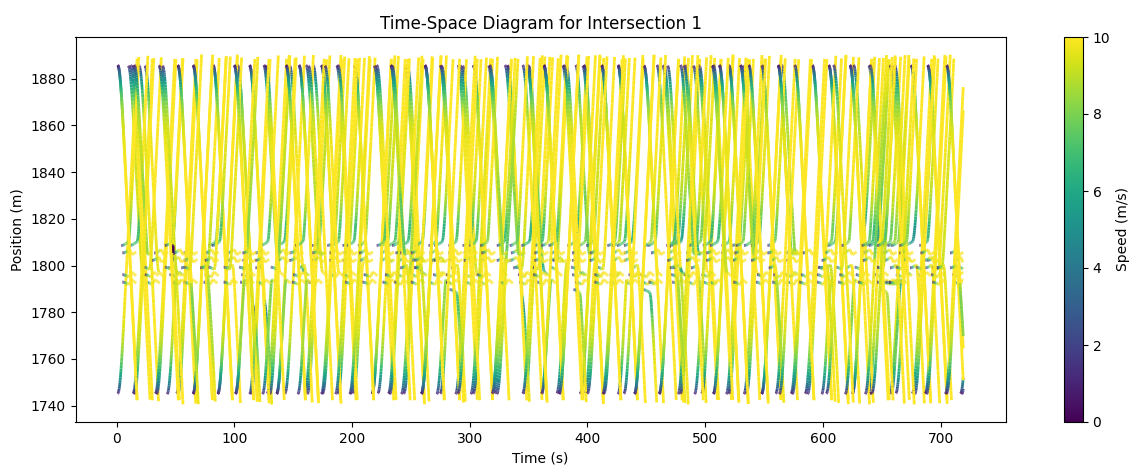}
   \caption{Time-space diagram (dataset 1)}
   \label{fig:time_space_dataset_1}
\end{subfigure}\hspace{2pt} 
\begin{subfigure}{0.24\textwidth}
   \includegraphics[width=\linewidth, height=4cm]{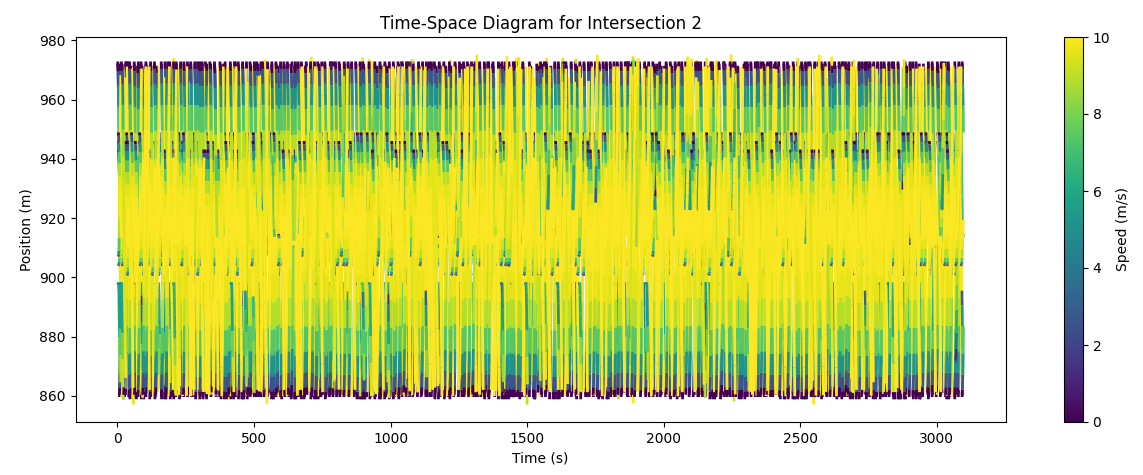}
   \caption{Time-space diagram (dataset 2)}
   \label{fig:time_space_dataset_2}
\end{subfigure}\hspace{2pt} 
\begin{subfigure}{0.24\textwidth}
   \includegraphics[width=\linewidth, height=4cm]{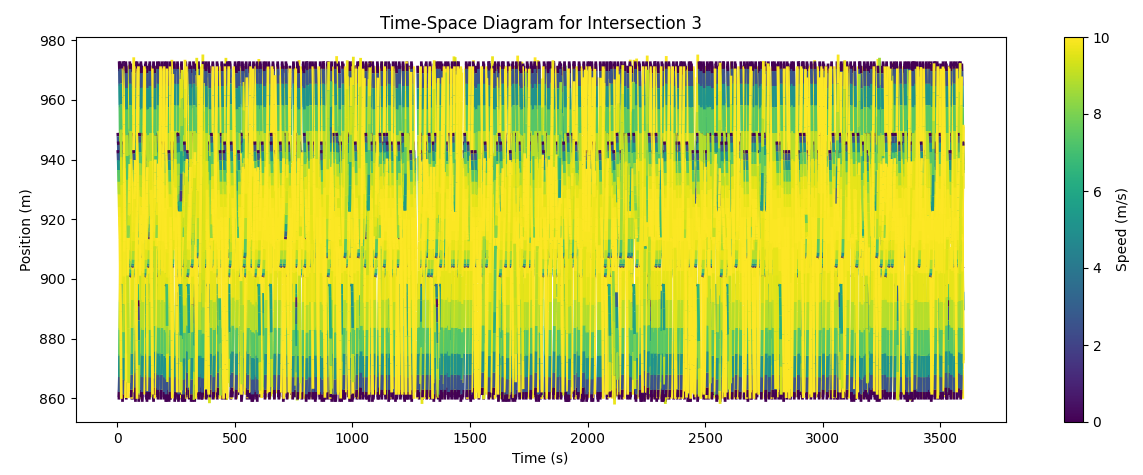}
   \caption{Time-Space diagram (dataset 3)}
   \label{fig:time_space_dataset_3}
\end{subfigure}\hspace{2pt} 
\begin{subfigure}{0.24\textwidth}
   \includegraphics[width=\linewidth, height=4cm]{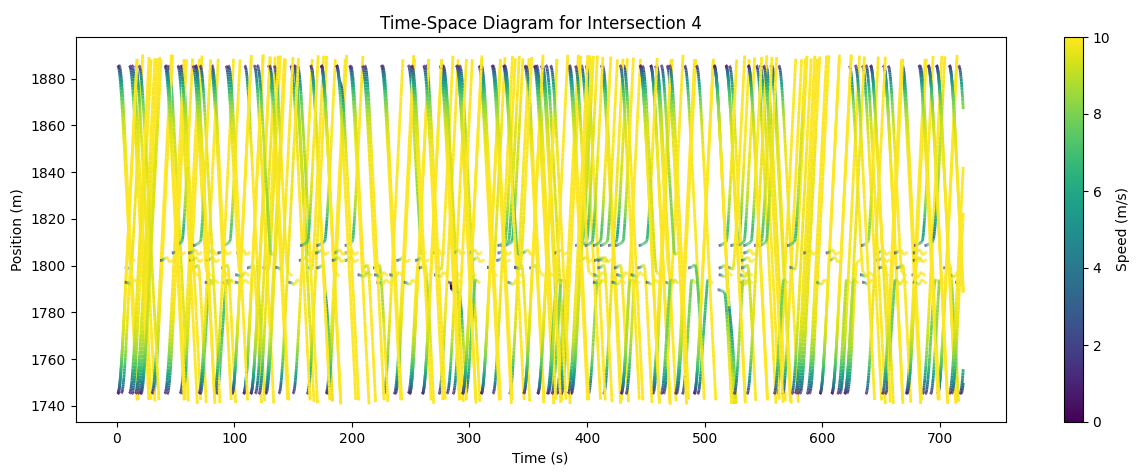}
   \caption{Time-space diagram (dataset 4)}
   \label{fig:time_space_dataset_4}
\end{subfigure}

\caption{Time-Space Diagram}
\label{fig:time_space}
\end{figure*}

In Fig.\ref{fig:time_space} x-axis represents time in seconds, while the y-axis represents the position along a stretch of road in meters. Each line in the diagram represents the trajectory of a single vehicle as it moves through space over time. The color of the lines corresponds to the vehicle speeds, with the color bar on the right side indicating speeds from 0 to 10 meters per second. Warmer colors (yellow) denote higher speeds, while cooler colors (blues and purples) indicate slower speeds or stops. The slope of a line indicates the vehicle's speed — a steeper slope suggests higher speed, while a flatter slope indicates that the vehicle is moving more slowly.

\begin{itemize}
    \item \textbf{Traffic Dynamics:} \\
    These time-space diagrams present some significant traffic dynamics such as congestion points, acceleration and deceleration, traffic waves, and so on. The horizontal stretches on the plot show where vehicles have either stopped or are moving very slowly. This could be due to congestion, a natural bottleneck at the intersection, or vehicles yielding to other traffic. We can observe in \ref{fig:time_space_dataset_1}, \ref{fig:time_space_dataset_2}, \ref{fig:time_space_dataset_3} and \ref{fig:time_space_dataset_4} that at certain times, lines are denser and change color towards the cooler end of the spectrum, indicating slower speeds which could suggest congestion or a slowdown in traffic flow. \\
    The steeper lines indicate periods where vehicles are accelerating, and the more horizontal lines indicate deceleration or stopping. It appears there are cycles of acceleration and deceleration, which could be due to the intermittent flow of traffic in an unsignalized intersection, where vehicles must yield or stop due to other vehicles crossing. These are phenomena where vehicles slow down (the lines cluster together and turn blue) and then speed up (the lines spread apart and turn yellow), which is typical in congested traffic conditions. \\
    The alternating patterns of color and the wave-like formations of the lines indicate the presence of traffic waves where vehicles periodically slow down and speed up. This can be a characteristic of traffic flows, especially in an unsignalized context where the flow is not regulated by traffic signals. \\
    The diagram also shows some instances of stop-and-go movement, where vehicles come to a near stop (lines become horizontal and blue) and then accelerate (lines become steep and yellow).

    \item \textbf{Traffic Patterns:} \\
    Traffic Patterns reveals a nature of uniformity and platooning. The regularity of patterns over time suggests a steady input of vehicles into the system. The start and end of vehicle input are likely responsible for the periodic nature of the traffic flow. \\
    Groups of vehicles seem to be moving together in clusters or platoons. This platooning effect is common in traffic flows and can be exacerbated by the stop-and-go patterns at unsignalized intersections.

    \item \textbf{Traffic Flow Implications} \\
    The efficiency of the intersection can be inferred from the slope of the lines. A higher slope (indicating higher speed) means less time spent by vehicles in the system, which is desirable. 
    The variation in speed and the presence of stopping and starting behavior could have implications for safety, as these are conditions where accidents are more likely to occur.
\end{itemize}

\subsection{Travel Time Analysis in Intersection}

\begin{figure*}[htbp]
\centering
\begin{subfigure}{0.24\textwidth}\hspace{2pt}
   \includegraphics[width=\linewidth, height=4cm]{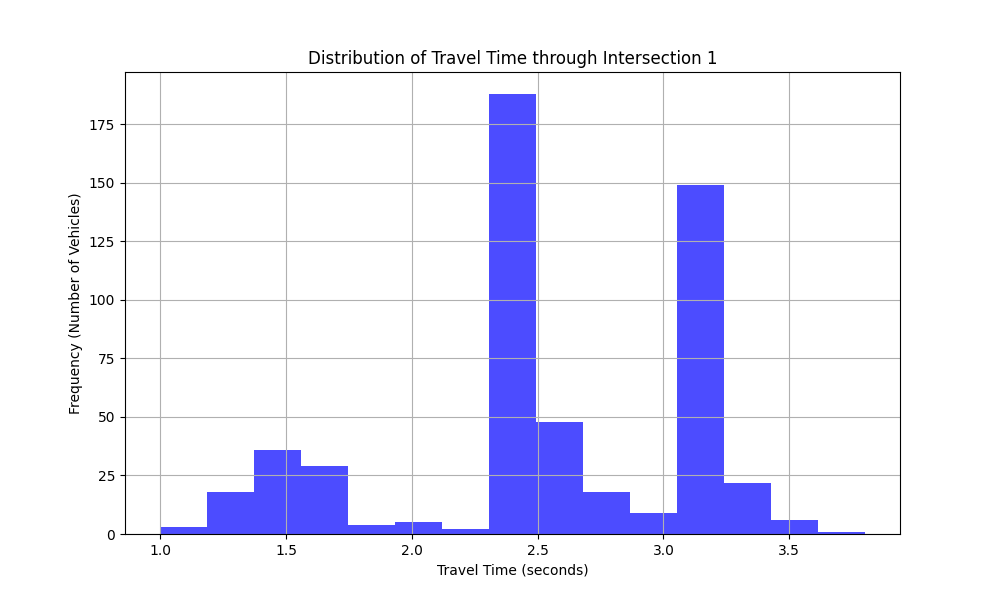}
\caption{dataset 1 histogram intersection travel time}
   \label{fig:intersection_2_histogram_intersection_travel_time}
\end{subfigure}%
\begin{subfigure}{0.24\textwidth}\hspace{2pt}
   \includegraphics[width=\linewidth, height=4cm]{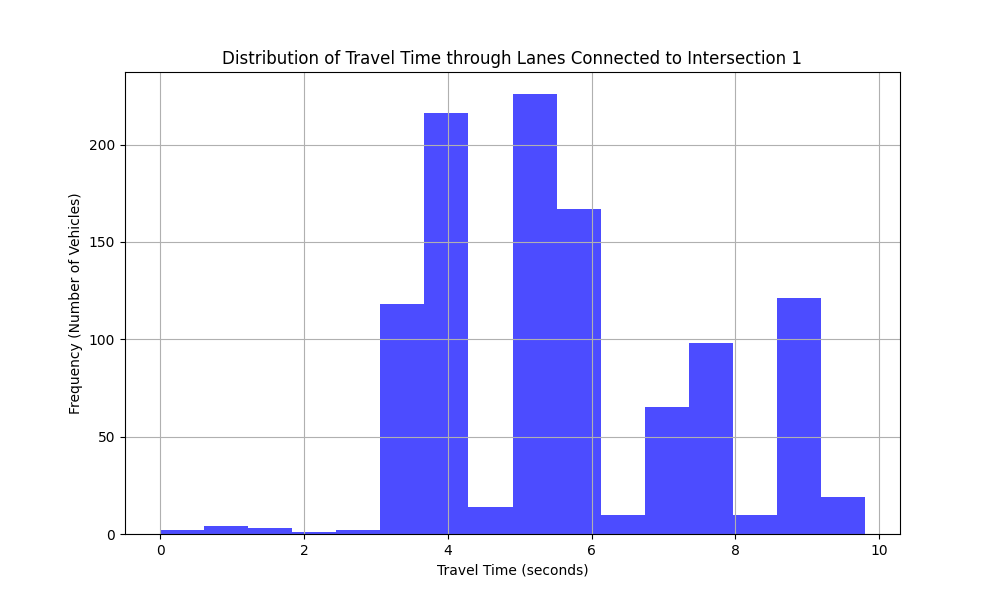}
   \caption{Dataset 1 histogram lanes travel time}
   \label{fig:intersection_2_histogram_lanes_travel_time}
\end{subfigure}%
\begin{subfigure}{0.24\textwidth}\hspace{2pt}
   \includegraphics[width=\linewidth, height=4cm]{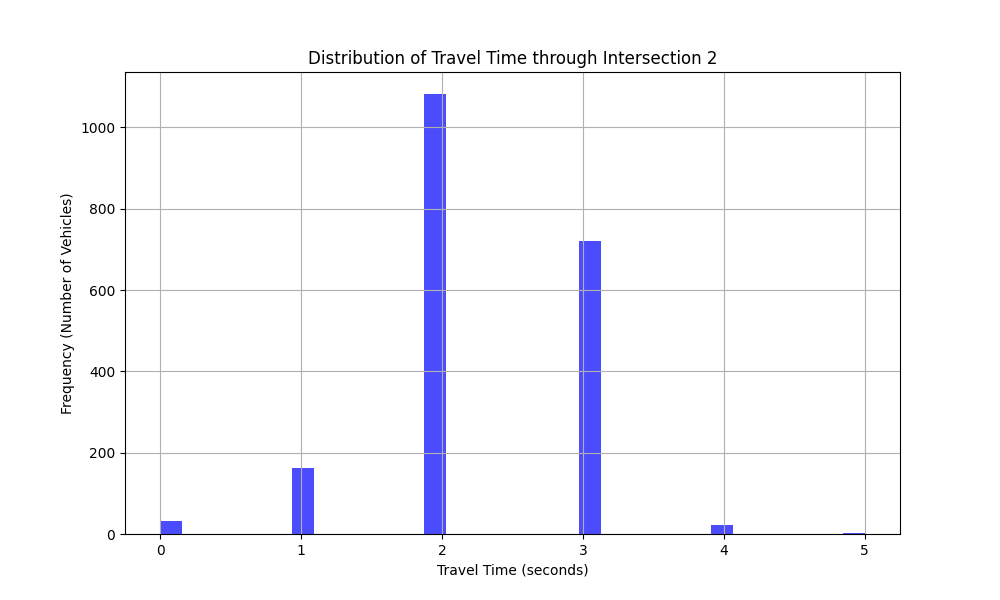}
   \caption{Dataset 2 histogram intersection travel time}
   \label{fig:midday_histogram_intersection_travel_time}
\end{subfigure}%
\begin{subfigure}{0.24\textwidth}\hspace{2pt}
   \includegraphics[width=\linewidth, height=4cm]{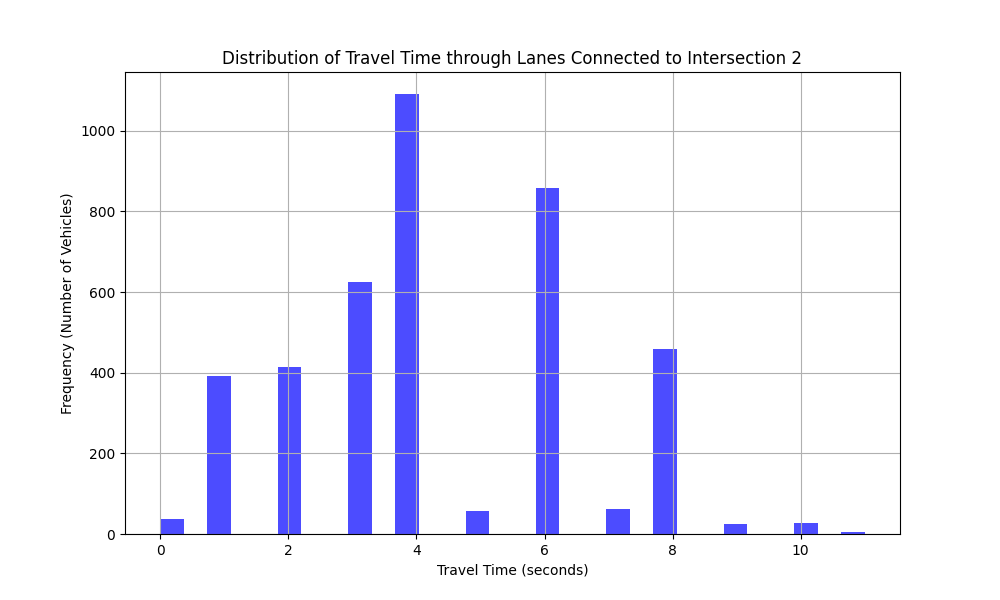}
   \caption{Dataset 2 histogram lanes travel time}
   \label{fig:midday_histogram_lanes_travel_time}
\end{subfigure}

\begin{subfigure}{0.24\textwidth}
   \includegraphics[width=\linewidth, height=4cm]{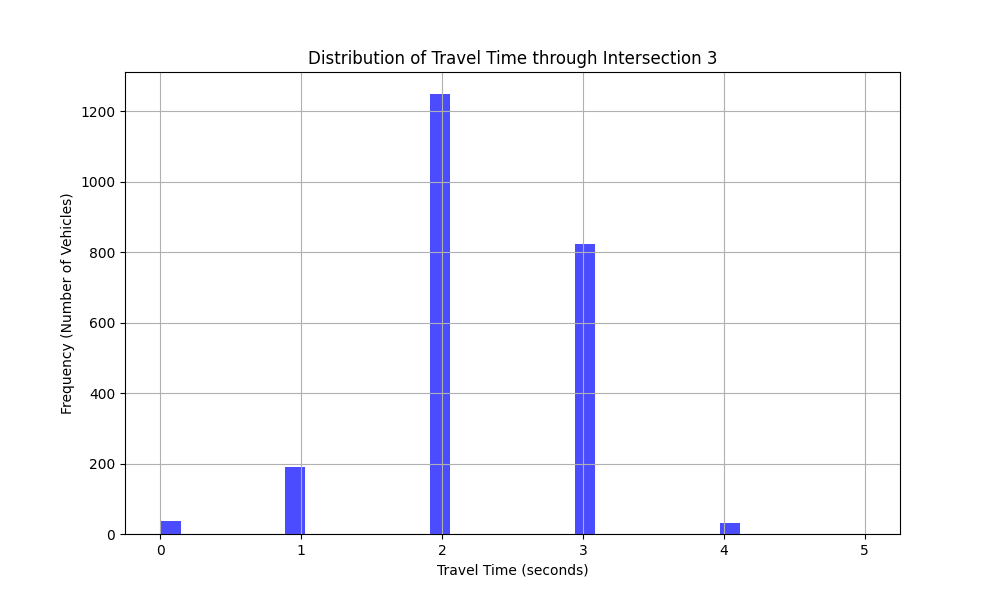}
   \caption{Dataset 3 histogram intersection travel time}
   \label{fig:rush_hour_histogram_intersection_travel_time}
\end{subfigure}%
\begin{subfigure}{0.24\textwidth}
   \includegraphics[width=\linewidth, height=4cm]{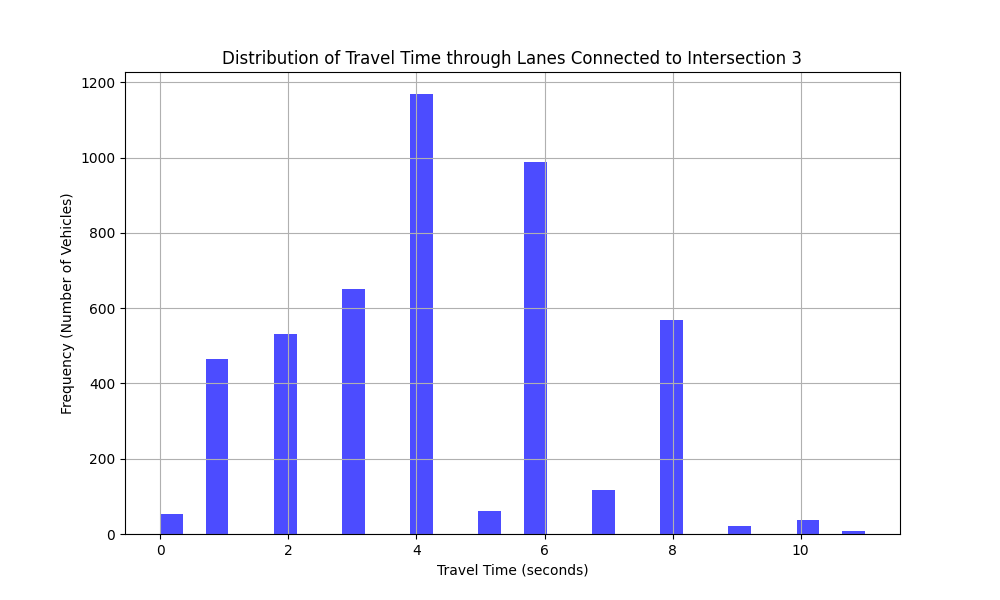}
   \caption{Dataset 3 histogram lanes travel time}
   \label{fig:rush_hour_histogram_lanes_travel_time}
\end{subfigure}%
\begin{subfigure}{0.24\textwidth}
   \includegraphics[width=\linewidth, height=4cm]{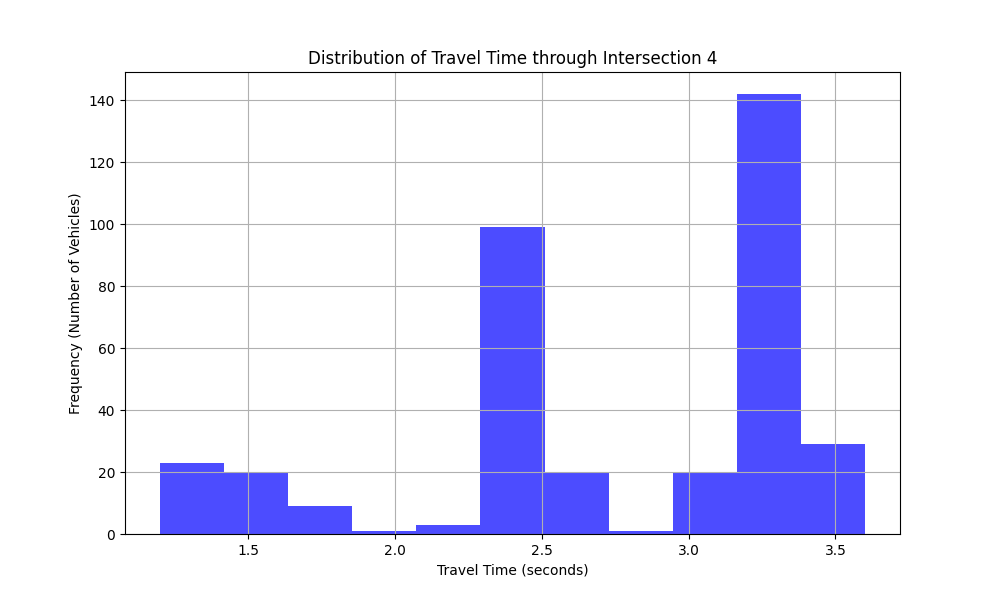}
   \caption{Dataset 4 histogram intersection travel time}
   \label{fig:lower-group2-image1}
\end{subfigure}%
\begin{subfigure}{0.24\textwidth}
   \includegraphics[width=\linewidth, height=4cm]{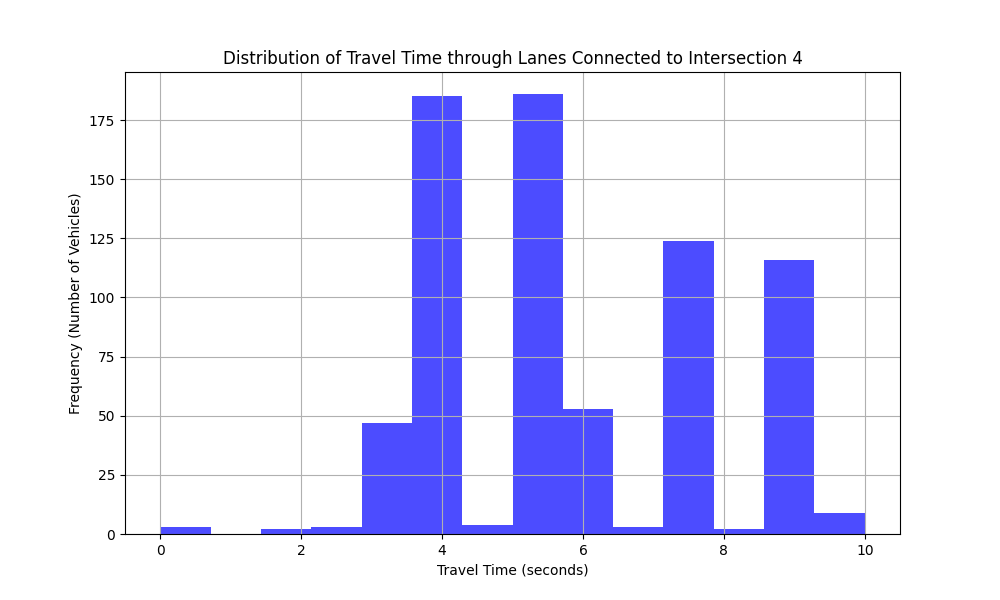}
   \caption{Dataset 4 histogram lanes travel time}
   \label{fig:intersection_4_histogram_lanes_travel_time}
\end{subfigure}

\caption{Frequency distribution of travel time}
\label{fig:histogram}
\end{figure*}

Figure~\ref{fig:histogram} offers a visual representation of travel times through four unsignalized intersection scenarios. X-axis axis quantifies the travel times, with each bin representing a range of travel times. Shorter travel times can be indicative of smooth traffic flow, while longer times may suggest delays. The Y-axis indicates how many vehicles fall within each travel time bin. Higher frequencies in certain bins signal that many vehicles are experiencing similar travel times. The tallest bars indicate the most common travel times. At an unsignalized intersection, these peaks may reflect natural gaps in traffic that allow vehicles to pass through without stopping or yielding. A widespread suggests variability, which in an unsignalized intersection, could be due to varying traffic volumes, differences in driver behavior, or the impact of crossing pedestrians or cyclists. Gaps or lower bars might show less frequent travel times, potentially reflecting variations in traffic flow or times when vehicles must yield to others. Multiple peaks could indicate different dominant traffic patterns, perhaps associated with varying levels of cross traffic or different times of day. A skew in the histogram might indicate a tendency towards either quicker or more delayed travel times. In unsignalized intersections, a right skew could suggest that while most vehicles pass through quickly, there is a tail of vehicles experiencing longer delays.

In Figure~\ref{fig:intersection_2_histogram_intersection_travel_time} each bin represents a range of travel times, and the width of these bins seems to be 0.1 seconds. Most vehicles have a travel time of around 1.0 to 1.5 seconds or 3.0 to 3.5 seconds, which are the two prominent peaks. There are fewer vehicles with travel times between these two ranges, indicating a bimodal distribution. The faster mode (1.0 to 1.5 seconds) could represent vehicles that do not need to stop or yield, possibly because they have the right-of-way or because the intersection is clear when they approach. The slower mode (3.0 to 3.5 seconds) may reflect vehicles that need to slow down, yield to other traffic, or stop before proceeding. The distribution suggests a relatively well-functioning give-and-take system where vehicles either pass quickly through the intersection or take longer due to waiting for their turn to proceed.
The lack of vehicles with intermediate travel times could mean that most vehicles either find a clear intersection or must yield to others, with little in-between. The presence of two peaks indicates that while many vehicles can pass through the intersection with little delay, a significant number also experience longer travel times, which could indicate waiting for a gap in traffic. The variation in travel times could also indicate points where conflict or hesitation is occurring.  

Given the unsignalized nature of the intersection, the presence of two prominent peaks in your histogram might suggest that there are two prevalent travel time experiences. This could be due to factors such as:

Intermittent High Traffic Volumes: Vehicles may arrive in platoons, leading to periods of higher and lower travel times.
Yielding Behavior: Drivers may have to yield to other vehicles, leading to variable travel times.
Cross Traffic: The amount and timing of cross traffic can significantly impact travel times at an unsignalized intersection.
Analyzing this distribution helps in understanding the efficiency of the intersection's design. For example, if a significant number of vehicles experience longer travel times, this might indicate that the intersection design is not optimal and could benefit from improvements. This could include physical changes to the intersection layout, the addition of traffic control measures, or other interventions to promote a more consistent and safer flow of traffic.

In \ref{fig:intersection_4_histogram_lanes_travel_time}, the histogram has a notable peak for travel times between 3.0 to 3.5 seconds. There are smaller peaks at around 1.5 seconds and just above 2.0 seconds. The distribution is right-skewed, with a tail extending towards longer travel times. The most frequent travel time is between 3.0 to 3.5 seconds, considering it as an unsignalized intersection scenario, the primary peak at 3.0 to 3.5 seconds might represent vehicles slowing down to yield to others or to ensure it’s safe to proceed. Smaller peaks indicate that some vehicles are passing through the intersection more quickly, possibly due to less traffic or more straightforward maneuvers such as not having to yield to others. The right-skewed distribution suggests that while the majority of vehicles pass through within a certain time frame, a considerable number experience delays, potentially due to waiting for a safe gap in traffic or negotiating with other drivers. The variability and spread of the histogram could also imply different levels of caution or hesitation among drivers at the intersection. The delays represented by the tail of the histogram could be due to increased traffic volume, vehicles making turns, or other factors that require drivers to slow down or stop. 

In Figure \ref{fig:midday_histogram_intersection_travel_time}, there are significant peaks at 2 seconds and 3 seconds. These peaks are very high compared to the other bins, indicating a large number of vehicles passing through the intersection within these times. The histogram shows a very low frequency for other travel times, which are near-zero for times other than the peaks. The peak at 2 seconds suggests that many vehicles are able to traverse the intersection quickly, possibly indicating a clear path or a priority movement that does not require stopping or yielding. The peak at 3 seconds indicates another common travel time, which may represent vehicles that slow down slightly more, perhaps due to yielding or minor delays. The sharp peaks at specific times with little to no vehicles in between suggest that the intersection may have very specific conditions or rules that cause vehicles to pass through at these distinct times. The distribution could reflect the presence of a dominant traffic flow or direction that allows vehicles to pass with minimal delay, with a secondary flow requiring a slight wait. Traffic engineers might use this kind of data to understand the efficiency of an intersection and to see if there are opportunities to improve flow.
Given the right-skewed distribution, with fewer vehicles taking longer than 3 seconds, there may be opportunities to investigate what causes the delays for these vehicles.

Figure~\ref{fig:rush_hour_histogram_intersection_travel_time} have travel time bins set at 1-second intervals from 0 to 5 seconds. There are two prominent peaks at 2 seconds and 3 seconds, where a high number of vehicles have these travel times. The frequencies at 1 second, 4 seconds, and 5 seconds are very low in comparison. The peak at 2 seconds could suggest that a substantial portion of vehicles pass through the intersection relatively quickly, perhaps because there is little cross-traffic or no requirement to yield. The peak at 3 seconds indicates another common travel time, which could be due to vehicles taking a bit more time to navigate the intersection, possibly due to yielding to other traffic or minor delays. The fact that most vehicles are concentrated in these two specific travel times implies that the intersection has a dominant traffic flow, and vehicles adhere to a predictable pattern of movement.
The near absence of vehicles in the other bins suggests that there is little variability in how vehicles pass through the intersection, possibly due to consistent traffic conditions or behavior patterns. For an unsignalized intersection, the distribution could indicate well-established right-of-way rules or patterns where drivers are accustomed to the flow and can navigate the intersection with relative predictability.

Intersection 1 had two clear modes of travel time, while Intersections 2 and 3 also showed bimodality but with more vehicles concentrated in the higher travel time peak. Intersection 4 showed a less defined pattern with a broader range of travel times. Intersection 1 had evenly distributed peaks, Intersections 2 and 3 had a dominant peak at 3 seconds, and Intersection 4's primary peak was also around 3 seconds but with additional minor peaks. Intersection 1 had a more uniform distribution between its two peaks. In contrast, Intersections 2 and 3 had sharper peaks with fewer vehicles traveling at times between the peaks. Intersection 4 showed a more gradual distribution. Intersections 2 and 3 displayed more consistency in travel times with the sharp peaks, whereas Intersection 1 and particularly Intersection 4 showed more variability.
\subsection{Traffic Flow Analysis}
The purpose of plotting average speed against position in the intersection lanes is to determine potential bottlenecks and analyze the flow of traffic through the intersections.

\begin{figure*}[htbp]
\centering
\begin{subfigure}{0.24\textwidth}
   \includegraphics[width=\linewidth, height=4cm]{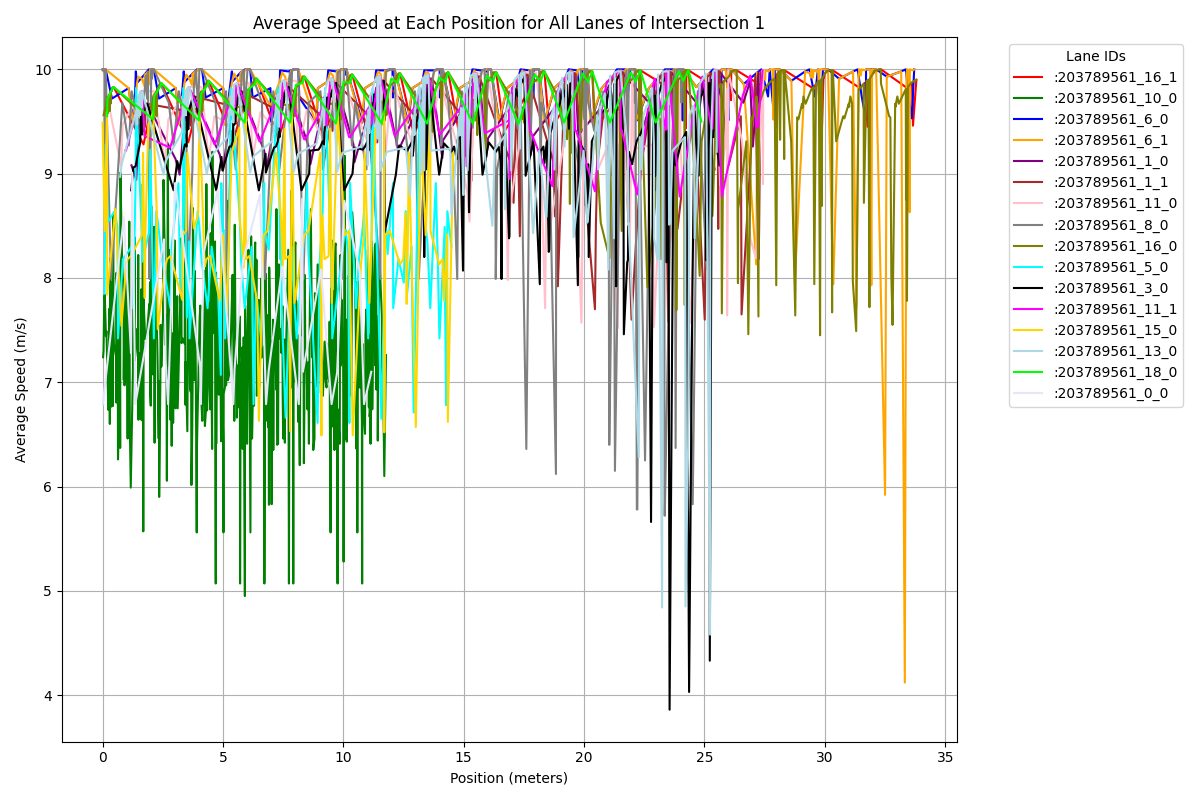}
\caption{Dataset 1 position vs speed inside intersection lanes}
   \label{fig:intersection_2_pos_speed_lanes}
\end{subfigure}%
\begin{subfigure}{0.24\textwidth}
   \includegraphics[width=\linewidth, height=4cm]{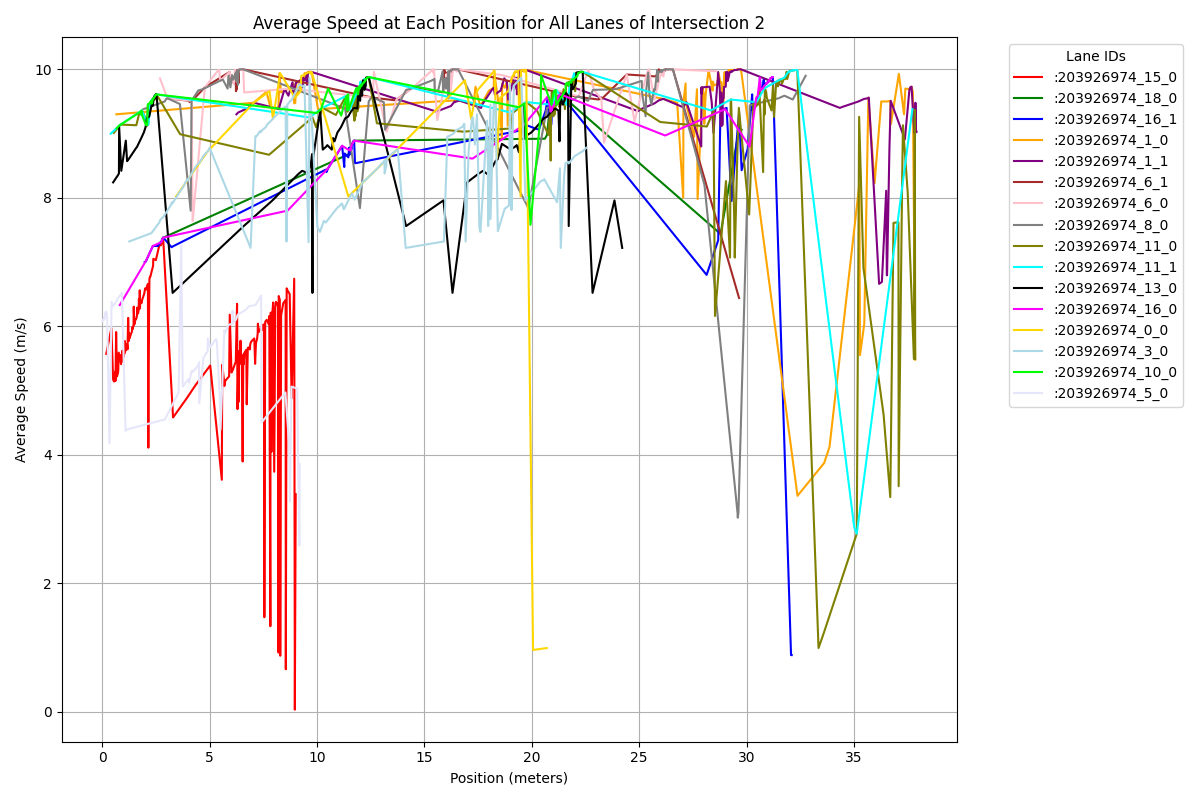}
   \caption{Dataset 2 pos vs speed lanes}
   \label{fig:midday_pos_speed_lanes}
\end{subfigure}%
\begin{subfigure}{0.24\textwidth}
   \includegraphics[width=\linewidth, height=4cm]{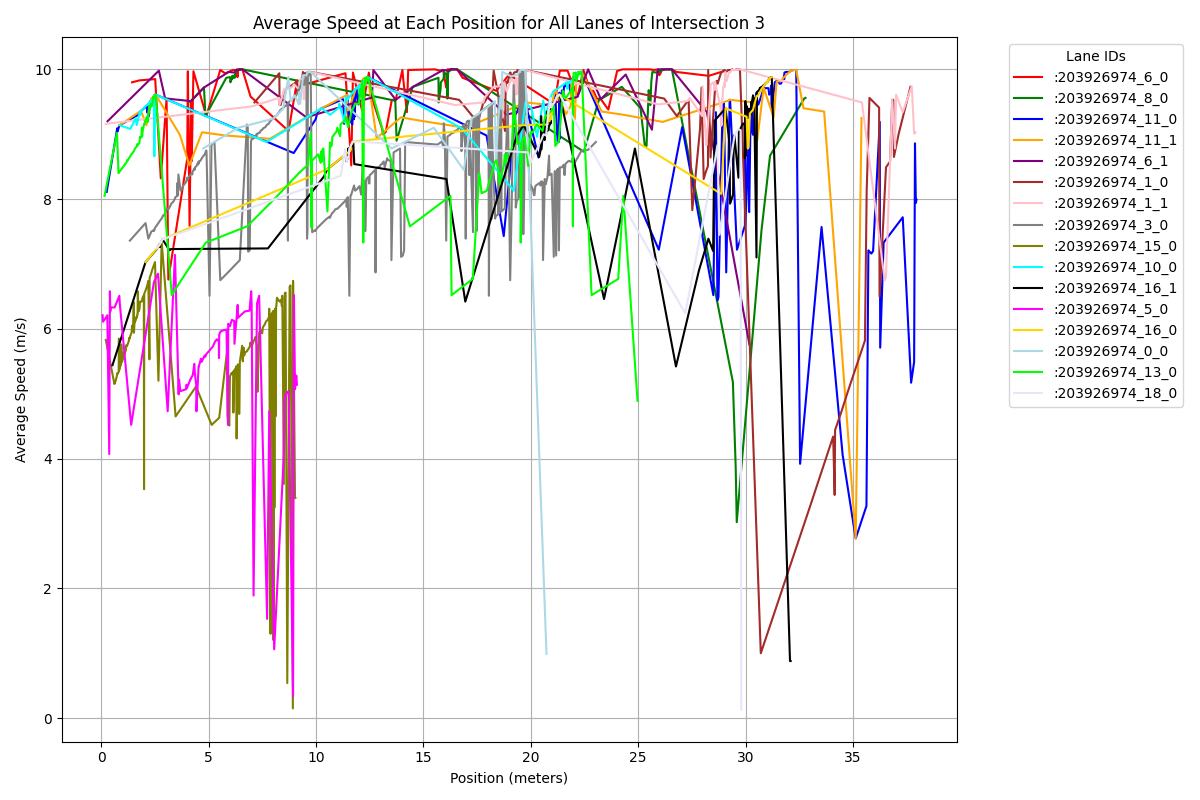}
   \caption{Dataset 3 pos vs speed lanes}
   \label{fig:rush_hour_pos_speed_lanes}
\end{subfigure}%
\begin{subfigure}{0.24\textwidth}\hspace{2pt}
   \includegraphics[width=\linewidth, height=4cm]{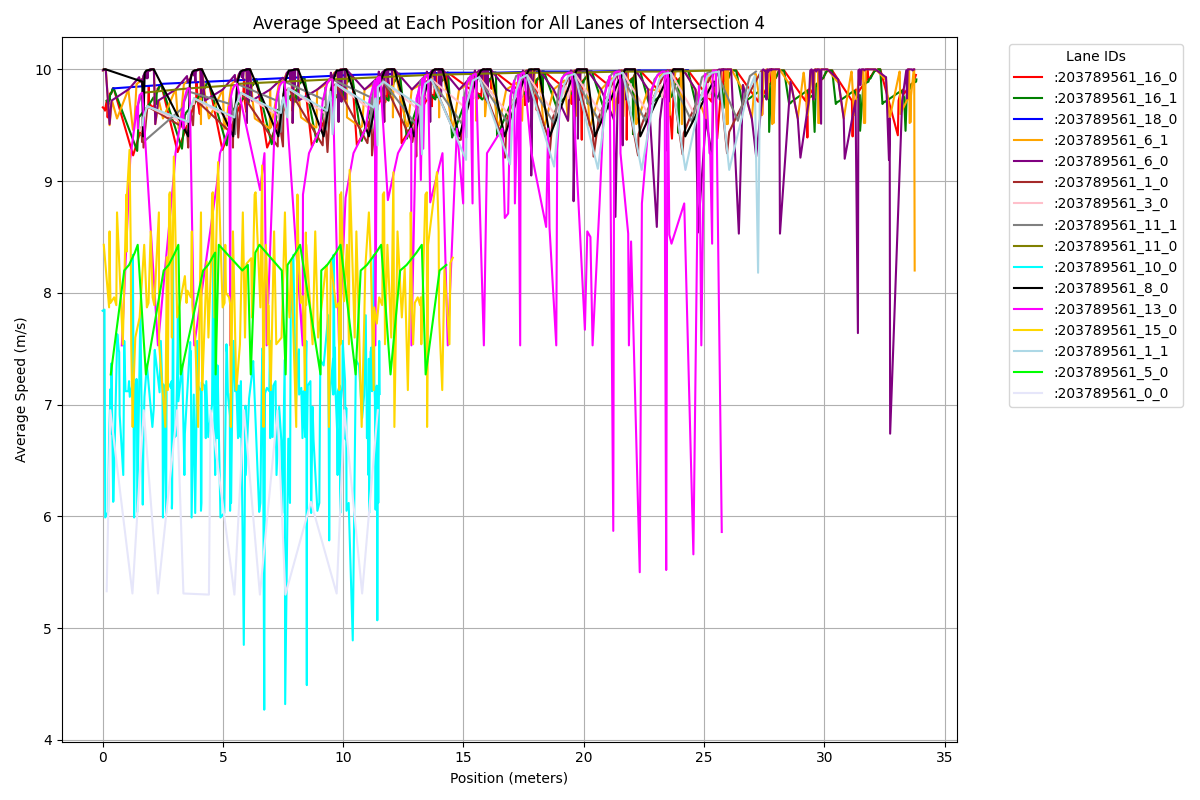}
   \caption{Dataset 4 pos vs speed lanes}
   \label{fig:intersection_4_pos_speed_lanes}
\end{subfigure}


\caption{Average speed in different position of lanes inside intersections for dataset 1, dataset 2, dataset 3 and dataset 4}
\label{fig:avg_speed_positions}
\end{figure*}

In \ref{fig:avg_speed_positions}, the x-axis represents the position in meters, which could correspond to different points along the approach or within the intersection itself. The y-axis represents the average speed of vehicles in meters per second (m/s). Each line represents the average speed profile of vehicles in a specific lane.
There's considerable variability between the lanes. Some lanes show speed patterns that increase, decrease, or fluctuate significantly, indicating varying traffic conditions or driver behaviors. Several lanes show a sharp decrease in speed at certain positions, which could indicate a common point where vehicles are required to slow down or stop, possibly due to a stop sign, yield sign, pedestrian crossing, or congestion.
Some lanes maintain a relatively stable speed throughout, suggesting a smoother flow of traffic in those lanes. 
The graph shows that speeds fluctuate greatly, with some lanes experiencing sharp drops to near 0 m/s. This could indicate stopping or yielding behavior at the intersection.
There are also lanes where the speed increases after a drop, which might indicate acceleration after stopping or yielding.

The plot provided is a complex visualization showing the average speed of vehicles at various positions for multiple lanes within an intersection, with each lane represented by a different color. Here's how to analyze the plot for potential bottlenecks:

Variability Across Lanes: Look for lanes where the average speed significantly drops, which could indicate a bottleneck. These are typically characterized by sharp and sustained dips in the graph.

Common Trends: If multiple lanes show a decrease in speed at similar positions, this might point to an intersection-wide issue affecting several lanes, such as a poorly designed junction or an obstacle affecting multiple paths.

Individual Lane Analysis: Examine each lane's graph separately:

Sustained Low Speeds: Prolonged sections of the graph with lower speeds suggest congestion or a consistent slowdown factor.
Sharp Decreases in Speed: Points where the speed sharply drops could be where vehicles are required to stop or yield, such as crosswalks, stop signs, or areas with high pedestrian activity.
Comparative Speed Drops: Compare the speed profiles between lanes. A lane with distinctly lower speeds compared to others at the same positions might have specific issues, like narrower widths, sharper turns, or other restrictive conditions.

From the plot, potential bottlenecks are likely to be found at positions where the lines consistently reach the lowest points across multiple lanes. For a more detailed analysis, you could:

Zoom in on Specific Lanes: By isolating each lane in a separate plot, you can better identify precise points of speed reduction.
Overlay Physical Map: Overlaying these data on a physical map of the intersection can help correlate speed drops with specific intersection features.
Cross-Reference with Incident Reports: Check if areas of low speed correspond with known sites of frequent traffic incidents or complaints.
If you wish to focus on a particular lane, you can trace the line corresponding to that lane's color in the legend and follow it across the plot. Look for patterns of speed reduction and consider if they are unique to that lane or part of a larger pattern affecting the intersection.

In summary, for lanes where the graph shows consistently low speeds or significant drops in speed, further investigation into the causes of these slowdowns is warranted. This might involve examining the intersection layout, timing of any traffic control devices, or the behavior of traffic at different times of day.
\subsubsection{Identify Peak Times}

\subsection{Discussion} 
\subsubsection{Comparative analysis of 4 intersection scenarios}

\begin{itemize}
    \item In the first image (Intersection scenario 4), the traffic flow seems to have more instances of slower speeds or stops, as indicated by the purple areas and horizontal lines, particularly in the middle of the graph.
    \item In the \ref{fig:time_space_dataset_1} (Intersection scenario 1), the overall color is more consistently green and yellow, suggesting that vehicles maintain a higher speed through the intersection.
    \item These observations suggest that during the time represented by the intersection scenario 1, the intersection experienced more stop-and-go behavior, which could indicate heavier traffic, potential obstructions, or other factors that caused vehicles to slow down or stop more frequently. On the other hand, the intersection scenario 4 shows a smoother flow of traffic with vehicles maintaining higher speeds.
    \item By comparing the patterns and colors shown in Fig \ref{fig:time_space_dataset_1} and \ref{fig:time_space_dataset_4}, we can conclude that the traffic conditions at the intersection varied between the two times of day, with the intersection scenario 4 showing a generally faster and smoother flow of traffic. 

    \item Potential for Improvements for 1: \\
    The data could be used to assess whether a change in control at the intersection might improve flow or safety—such as adding a roundabout, or other traffic calming measures. If the slower travel times are due to vehicles waiting for gaps in busy cross traffic, a traffic control measure might be warranted to improve efficiency and safety.
\end{itemize}

This kind of analysis is crucial for understanding traffic behavior and can inform decisions on how to improve the flow of traffic in an unsignalized traffic environment.

\section{CONCLUSIONS}
\label{conclusion}

This research has successfully demonstrated the value of integrating manually labeled video data with advanced simulation tools like SUMO to analyze traffic dynamics at unsignalized intersections. Our comprehensive analysis, using techniques like time-space diagrams, travel time distribution, and speed-position correlations, has unveiled critical insights into traffic flow and bottleneck points. These findings have significant implications for traffic management and infrastructure planning, particularly in urban areas like Memphis, TN. This approach underscores the potential for detailed traffic scenario reconstruction in enhancing our understanding of complex urban traffic systems and guiding effective urban planning and policy decisions.

There are many future research directions. 
First, we aim to analyze the RVs at the trajectory level~\cite{Li2019ADAPS,Shen2022IRL,Lin2022Attention}, under adversarial conditions~\cite{Poudel2021TrafficStateAttack,Shen2021Corruption,Villarreal2022AutoJoin}, and possibly with hardware~\cite{Poudel2022Micro}. This analysis intends to determine whether combining different elements would enhance the capabilities of robot vehicles further.
Secondly, we aim to scale up our study by incorporating intersectional mixed traffic into larger areas, ideally citywide road network. This will involve leveraging existing techniques for large-scale traffic simulation, reconstruction, and prediction~\cite{Wilkie2015Virtual,Li2017CityFlowRecon,Li2017CitySparseITSM,Li2018CityEstIET,Lin2019Compress,Lin2019BikeTRB,Chao2020Survey,Lin2022GCGRNN}. 
Third, we want to analyze the impact of various vehicle networks via network optimization~\cite{Wickman2022SparRL}, and heterogeneous traffic via multi-agent learning and optimization~\cite{Wickman2023Species,wickman2021lrn}. 
Lastly, we want to combine mixed traffic modeling with crowd simulation and virtual humans~\cite{Li2013Memory,Durupinar2016Individual,Li2012Commonsense,Li2012Apprentice,Li2012Distribution,Li2011Purpose} so that together we can model more realistic road uses with pedestrians as intelligent virtual characters.







\bibliographystyle{unsrt}
\bibliography{ref}

\end{document}